\documentclass[12pt]{article}
\usepackage{latexsym}

\def\cH{{\cal H}}

\def\cS{{\cal S}}
\def\cD{{\cal D}}

\def\tr{{\rm tr}}
\def\ket#1{\mid~\!\!\!{#1}~\!\!\rangle}
\def\bra#1{\langle~\!\!{#1}~\!\!\!\mid}

\def\N{_{1\dots N}}

\def\6{\Big[(N!)\Big/\prod_{j=1}^J(N_j!)
\Big]}

\def\M{_{(M_j+1)\dots (M_j+N_j)}}

\def\QM{{\rm quantum mechanics }}
\def\qm{{\rm quantum mechanics}}
\def\Q{quantum }

\def\QMl{\rm quantum-mechanical }

\def\${\enspace$}

\begin{document}

\begin{center} {\LARGE \bf On the nucleon paradigm:\\ the nucleons are closer to reality\\ than the protons and neutrons} \vspace{.5cm}

\bf \large  Fedor Herbut\\
\end{center}

\noindent {\footnotesize \it Serbian Academy
of Sciences and Arts, Serbia, Belgrade, Knez
Mihajlova 35}

\vspace{0.5cm} \noindent \rule{13.6cm}{.4pt}

\noindent {\bf Abstract}\\
\indent There is a widespread delusion that in theoretical nuclear physics protons and neutrons are the real thing, and nucleons are not more than a mathematically equivalent formality. It is shown that, on the contrary, nucleons are the real thing, because only a part of the theory is essentially identical to proton-and-neutron theory, whereas the remaining part is physically relevant. The approach is general. Thus, this is a paradigm of relation of a wider and a more narrow theory, so that the wider theory describes reality better. Also the relation of disjoint domains to the exclusion principle is clarified. A general fermion theory of how to distinguish identical particles is presented.\\

\noindent \rule{13.6cm}{.4pt}

\vspace{0.5cm} \normalsize \rm

\section{Introduction}

\indent
Nuclei with electron shells make up atoms and molecules, and further all the objects of the world of classical physics that we are familiar with.

What do nuclei consist of, are they protons and neutrons, or nucleons? I believe that many physicists would readily opt for the former. Some would choose the latter.

Protons and neutrons differ in mass, electric charge, and in gyro-magnetic coefficient. It is hard to believe in the reality of one, apparently fictitious particle, the nucleon, of which the proton and the neutron are two izospin-
\rule{2cm}{.4pt}

\indent {\scriptsize {\it E-mail address:}
fedorh@sanu.ac.rs}
\pagebreak

\noindent
projection states, analogous to spin up and spin down respectively. One can establish formal equivalence, but reality is another issue. Along these lines argue the believers in the reality of protons and neutrons.

But, the plausibility of such a line of reasoning rests on intuition that we have acquired in the school of thought of classical physics. Quantum physics teaches us new intuition. It is based more on formal, but relevant and precise, ideas than on simple and concrete, classically picturable notions. This may be a starting standpoint of the believers in the reality of nucleons.

Putting the dilemma in precise terms, there are two possibilities:

(i) Nucleons are real. Protons and neutrons have restricted reality.

(ii) Protons and neutrons are real; nucleons are formally, in a theoretical sense, equivalent to the former.

Surely, the reply may, as a prerequisite, demand taking stand in the following quest in natural philosophy:

(a) Quantum physics describes reality, 'existing out there', striving to do it as well as possible.

(b) 'Anti-realism' rejects (a), and reduces quantum phenomena to the unquestionable reality of classical physics plus \Q formalism.

(c) 'A-realism' does not care (Adler, 1989). (As some say, "shut up and calculate".)

The present author decidedly takes (a) as his standpoint. But still, besides this ontological point of view, there is an epistemological question: How well can we describe reality? We must not be deluded that we have any theory that corresponds to reality in an exact manner. Instead, as mathematicians would say, we can only approximately describe reality, and we strive to improve our approximation. The well-known and very fundamental irrational \$\pi =3.14...\$ is a good paradigmatic notion. One knows how to improve on the approximation, but not how to write down the precise value.

Two points follow from the relation of ontology and epistemology. Firstly, a theoretical description appears also in versions (b) and (c). In (b) it is all that there is in \Q physics (a rather poor ontology); in (c) one couldn't care less about ontology.

Secondly, one must clarify how one expects to {\bf 'improve'} one's description of reality.

My answer is that a 'better' theory must be {\bf wider}, and, in some sense, it must {\bf 'contain'} the former theory, and it must give {\bf 'more'}. The latter must be {\bf physically meaningful}.

In this article "the nucleon paradigm" (from the title) is conceived as a theory of nucleons in which it is precisely defined how it is wider and how it contains the description of protons and neutrons, and it is shown what it gives 'more' then a proton-neutron theory in a physically meaningful way.\\

Actually, this article has a greater ambition. It will give answer to another important question that concerns the exclusion principle of Pauli. One wonders whether one should, perhaps,

(i) anti-symmetrize all nucleons in the universe since they are all identical particles of the same kind;

(ii) or, one should anti-symmetrize only the nucleons that are as close as those in one and the same nucleus.

In case of (i), one must be able to show that though one anti-symmetrizes all nucleons in the universe, when relevant spatial domains are considered, it boils down to anti-symmetrizing special clusters of nucleons separately. In this way then, version (i) would 'contain' version (ii). The 'more' would come from the fact that the Pauli principle is based only on the identity of the particles and their spin without intrusion of spatial concepts as in (ii).

The theoretical framework that is going to be presented covers both cases (i) for the nucleons and case (i) for their anti-symmetrization. It will be potentially valid for any kind of identical particles, fermions or bosons.\\

The exposition of the 'wider', and hence closer to reality, theory in both mentioned cases will draw on a general theory of 'distinguishing identical fermions' explained in Appendices A and B.\\

The nucleon paradigm of {\bf two} nucleons was discussed in previous work (Herbut 2001). If the reader  finds the jump to the general case in this article, particularly Appendix B, to steep, he (or she) is well advised to reed the mentioned previous article first. But it treats only a very special case of the present theory.\\

\section{\bf The nucleon paradigm}
\indent Let a few remarks give a short {\it historical outline} of thinking that has led to the nucleon paradigm.

Following Jauch (1966), one
can distinguish {\bf intrinsic and
extrinsic properties} of
particles. According to him,
{\bf identical are those particles that
have equal intrinsic properties}.

Jauch's criterion seems to suggest that, to obtain a wider theory, one should treat some intrinsic properties of the particle as extrinsic. But this would be in vain unless it had a surplus of physical meaning.

De Muynck (1975) remarks on Jauch's criterion, "an intrinsic
property may show up dynamical
behavior", and turn out to be
extrinsic like the proton and
neutron states.

It all depends on the
experimental conditions. Mirman (1973) made the important claim that distinguishability of identical
particles is essentially an
experimental notion.\\

Let us resort to \QMl {\it state spaces} (complex Hilbert spaces). They are the natural mathematical framework for quantum state vectors (elements of the space of unit norm) or, more popularly, wave functions (state vectors in coordinate representation). They are \Q pure states (as opposed to mixed states).

The {\bf single-nucleon state
space} \$\cH_{nu}\$ has {\it three tensor-factor
spaces}: the orbital (or spatial)
one \$\cH_{orb}\$, which is a countably infinitely dimensional complex or a separable Hilbert space , the two-dimensional spin one \$\cH_{sp}\$, and the two-dimensional isospin one \$\cH_{iso}\$: $$\cH_{nu} \equiv \cH_{orb}\otimes\cH_{sp}\otimes\cH_{iso}.\eqno{(1a)}$$
This is so because the interactions between nucleons do not depend on the orientation of the z-component of spin, and, formally, the same is valid for the orientation of the third component \$t_3\$ of the isospin (so-called charge independence of the strong interaction).

Intuitively speaking, two protons, two neutrons or a proton and a neutron, act {\bf equally} as far as so-called {\bf strong interaction} is concerned. (This is not so for electromagnetic and weak interaction.) The proton is the counterpart of spin-up, i. e., it has \$t_3=1/2\$ (not in units of \$\bar h\$ ), whereas the neutron has \$t_3=-1/2\$.

Actually, the orbital and spin spaces in (1a)  are simultaneously also the state spaces of the {\bf proton} and the {\bf neutron}:
$$\cH_{pr}\equiv \cH_{orb}\otimes\cH_{sp},\eqno{(1b)}$$
$$\cH_{ne}\equiv \cH_{orb}\otimes\cH_{sp}.\eqno{(1c)}$$
The difference in the three \QMl state spaces (1a), (1b) an (1c) is in the {\bf intrinsic properties} of the particles, i. e., properties that {\it do not enter the \QMl formalism}, but {\it underly it}.

The nucleon is intrinsically only a barion (barion quantum number \$+1\$). Mass, electric charge and gyromagnetic factor are extrinsic properties. They enter the formalism; they depend on the state of the particle.

On the other hand, the proton has intrinsically the well-known mass, electric charge and gyromagnetic factor; whereas the neutron has, also intrinsically, its corresponding quantities.\\

It is customary in nuclear physics to denote the number of protons in a nucleus by Z, that of the neutrons in it by N, and their sum by A. To be consistent with our general notation in Appendix B, we write \$N_{pr}, N_{ne}\$ and \$N_{nu}\$ instead of \$Z, N\$ and \$A\$ respectively.

The {\bf \$N_{nu}$-identical-fermion state space} is $$\cH_{1\dots N_{nu}}^{nu}\equiv A_{1\dots N_{nu}}\prod_{n=1}^{\otimes, N_{nu}}\cH_n^{nu}\enskip ,\eqno{(2)}$$ where \$A_{1\dots N_{nu}}\$ denotes the anti-symmetrizer over all \$N_{nu}\$ particles (cf (A.2) in Appendix A). One should note that the index \$n\$ refers to the $n$-th particle in the (formal distinct-particle) $N_{nu}$-nucleon state space \$\prod_{n=1}^{\otimes, N_{nu}}\cH_n^{nu}\$. The superscript reminds of the intrinsic properties that define the identical particles.

One should note that the state space \$\cH_{1\dots N_{nu}}^{nu}\$ applies to the entire so-called {\bf isobar family} of nuclei, i. e., to all those nuclei that have the given number \$N_{nu}\$ of nucleons. It begins with \$N_{pr}=N_{nu}\$, \$N_{ne}=0\$, and it ends with \$N_{pr}=0\$, \$N_{ne}=N_{nu}\$. Besides these extreme cases, also many other members of the isobaric family usually do not exist in nature. (If obtained artificially, they, being unstable, decay with a certain half life.)\\

On the other hand, the proton-neutron description takes place in a {\bf one-nucleus proton-neutron state space} $$\cH^{pr,ne}_{1\dots N_{nu}}\equiv\cH^{pr}_{1\dots N_{pr}}\otimes
\cH^{ne}_{1\dots N_{ne}}\enskip ,\eqno{(3a)}$$ where $$\cH^{pr}_{1\dots N_{pr}}\equiv A_{1\dots N_{pr}}
\prod^{\otimes, N_{pr}}_{n=1}\cH_n^{pr}\enskip ,\eqno{(3b)}$$
$$\cH^{ne}_{1\dots N_{ne}}\equiv A_{1\dots N_{ne}}
\prod^{\otimes, N_{ne}}_{n=1}\cH_n^{ne}\enskip .\eqno{(3c)}$$ Here \$A_{1\dots N_{pr}}\$ and \$A_{1\dots N_{ne}}\$ denote the anti-symmetrizers, which are now applied separately to the two kinds of particles. (This is usually called the 'exclusion or Pauli principle', whereas the above nucleon anti-symmetrizer in (2) is called the 'extended exclusion' or 'Pauli principle'.) Finally, both the proton and the neutron have the same single orbital-spin space (cf (1b) and (1c)).\\

To see how the \$N_{nu}$-identical-fermion description in \$\cH_{1\dots N_{nu}}^{nu}\$ can be 'restricted' to the proton-neutron description in \$\cH\N^{pr,ne}\$, or how the former {\bf 'contains'} the latter (cf the Itroduction), {\bf distinguishing projectors} are required (cf Appendix B).

The {\bf single-nucleon
distinguishing projectors} in \$\cH_{nu}\$ (cf (1a)) are the
eigen-projectors \$Q_{nu}^+\$ and \$Q_{nu}^-\$ of \$t_3,\$ (the
third-projection of isospin by analogy with \$s_z\$ of spin). The spectral decomposition of \$t_3\$ is $$t_3=(1/2)Q_{nu}^++(-1/2)Q_{nu}^-.\eqno{(4)}$$ Actually,  all these operators act in the tensor-factor space  \$\cH_{iso}\$ (cf (1a)), and they are multiplied tensorically by the identity
operators in the orbital and in
the spin factor spaces to be determined in the entire single-nucleon state space \$\cH_{nu}\$. One uses the same notation in \$\cH_{iso}\$ and in \$\cH_{nu}\$. (One can see from the context in which space they act.)\\

To obtain the distinguishing projectors, with the physical meaning of the {\bf distinguishing property}, in the many-nucleon and the many-proton-neutron spaces, for a {\bf fixed nucleus}, i. e., with fixed numbers \$N_{pr}\$ and \$N_{ne}\$, the procedure goes as follows (cf Appendix B).

First we define the distinguishing many-particle projector in the $N_{nu}$-nucleon space \$\prod_{n=1}^{\otimes ,N_{nu}}\cH_n^{nu}\$ (cf (1a)).
$$Q_{1\dots N_{nu}}\equiv\Big( \prod_{n=1}^{\otimes ,N_{pr}}Q_n^+\Big)\otimes\Big(\prod_{n=N_{pr}+1}^{\otimes ,N_{nu}}Q_n^-\Big).\eqno{(5)}$$ In the extreme cases of no neutron or no proton, the formula, of course, requires slight obvious changes.

One should note that \$Q_{1\dots N_{nu}}\prod_{n=1}^{\otimes ,Nu}\cH_n^{nu}\$  is {\bf not} a subspace of the isobar family state space \$A_{1\dots N_{nu}}\prod_{n=1}^{\otimes ,Nu}\cH_n^{Nu}\$. It is a subspace of the formal distinct-particle space \$\prod_{n=1}^{\otimes ,Nu}\cH_n^{Nu}\$, and it is {\bf isomorphic} with the one-nucleus proton-neutron state space \$\cH^{pr}_{1\dots N_{pr}}\otimes
\cH^{ne}_{1\dots N_{ne}}\$ (cf (3a)). The isomorphism at issue is obtained by the transition from the one-nucleon space \$Q^+\cH_{nu}\$ (cf (1a)) to the proton state space \$\cH_{pr}\$ (cf (1b)) and from \$Q^-\cH_{nu}\$ to the neutron state space \$\cH_{ne}\$ (cf (1c)). This amounts to converting the relevant {\bf extrinsic properties} into {\bf intrinsic} ones (cf Jauch 1966).

In the {\bf isobar family state space} \$\cH_{1\dots N_{nu}}=A_{1\dots N_{u}}\prod_{n=1}^{\otimes ,N_{nu}}\cH_n^{nu}\$ one has the following symmetrized distinguishing projector (a sum of orthogonal projectors) that is the counterpart of \$Q\N\$ given by (5):

$$Q_{1\dots N_{nu}}^{sym}\equiv (N_{pr}!N_{ne}!)^{-1}\Big(\sum_{p\in S_{N_{nu}}}P_{1\dots N_{nu}}Q_{1\dots N_{nu}}(P_{1\dots N_{nu}})^{-1}\Big),\eqno{(6)}$$ where by \$S_N\$ is denoted, as customary, the symmetric group (or group of permutations) on \$N\$ objects.

The next step is to determine the subspace of the $N_{nu}$-identical-fermion state space \$\cH_{1\dots N_{Nu}}^{Nu}\$ (cf (2)) that is {\bf isomorphic} with the proton-neutron space of a {\bf fixed nucleus}.  It is (cf Theorem 1 and relation  (B.5)): $$\cH^{id}_{1\dots N_{nu}}\equiv Q_{1\dots N_{nu}}^{sym}A_{1\dots N_{nu}}\prod_{n=1}^{\otimes ,N_{nu}}\cH_n.\eqno{(7)}$$

As it was stated, the proton is the nucleon with \$t_3=1/2\$, and the neutron is the nucleon with \$t_3=-1/2\$. In the formalism this means that \$Q^+\cH_{nu}\$ and \$Q^-\cH_{nu}\$ are the proton and neutron state spaces respectively as subspaces of \$\cH_{nu}\$ (cf (1a)). To be more specific, \$Q^+\cH_{nu}\$ is isomorphic with the proton space \$\cH_{pr}\$ and  \$Q^-\cH_{nu}\$ with the neutron one \$\cH_{ne}\$ (cf (1a-c)). In practice, this isomorphism means, for both particles, to omit the isospin tensor factor space in (1a), and take over the orbital and spin spaces unchanged. Particularly, all orbital-spin operators remain unchanged.

The {\bf one-nucleus state space} in the proton-neutron description (3a), with the {\bf intrinsic} proton-or-neutron properties,  is not a subspace of the $N_{nu}$-nucleon distinct-particle state space \$\prod_{n=1}^{\otimes ,N_{nu}}\cH_n\$. But, if one rewrites (3a) in its {\bf isomorphic extrinsic form}, based on the insight of the preceding passage, as $$A_{1\dots N_{pr}}A_{(N_{pr}+1)\dots N_{nu}}Q_{1\dots N_{nu}}\prod_{n=1}^{\otimes ,N_{nu}}\cH_n^{nu}\eqno{(8)}$$ (cf (5)), then it is.

On ground of the general Theorem 1 in Appendix B, the one-nucleus state space \$\cH_{1\dots N_{nu}}^{id}\$ given by (7), which is a subspace of the all-nucleon space referring to the isobar family, and the one-nucleus state space (8) are {\bf isomorphic}, and the isomorphism {\bf acts as follows}: $$\Big[\Big(N_{nu}!\Big/N_{pr}!N_{ne}!\Big)^{1/2}
Q_{1\dots N_{nu}}\Big]\cH_{1\dots N_{nu}}^{id}=$$  $$A_{1\dots N_{pr}}A_{(N_{pr}+1)\dots N_{nu}}Q_{1\dots N_{nu}}\prod_{n=1}^{\otimes ,N_{nu}}\cH_n^{nu}.\eqno{(9)}$$\\

When {\bf weak interaction does not
play a role}, i. e., when no
\$\beta$-radioactivity is taking place, then the distinguishing property
\$Q\N^{sym}\$ is possessed by any \Q
state of the
nucleus. Namely, this property
physically simply says that there
are \$N_p\$ protons and \$N_n\$
neutrons in the \$N_{nu}$-nucleonic
nuclear state
(\$N_{nu}=N_p+N_n\$). Hence,
one can transfer the
quantum-mechanical description
from the first-principle
completely antisymmetric all-nucleon state space given by (2)
(in which the so-called 'extended Pauli principle' is valid) to the effective
distinct-cluster space given by (8), or even to the further isomorphic state space \$\cH\N^{pr,ne}\$ (cf (3a)). We have two clusters
here, to utilize the terminology of the general theory in Appendix B, that of protons and that of
neutrons.\\

When weak interaction (or
\$\beta$-radioactivity) has to be taken
into account, the single-particle
spaces \$\cH_{pr}\$ and \$\cH_{ne}\$ (relations (1b) and (1c) respectively)
have to be replaced by the doubly
dimensional {\it nucleon space} \$\cH_{nu}\$ given by (1a).\\

So-called weak interaction turns a neutron into a proton or {\it vice versa} within a nucleus. One observes this as $-\beta$ or $+\beta$ radioactivity (emission of an electron with a neutrino or emission of a positron with a corresponding neutrino) respectively. This displaces the nucleus in question to a neighboring one with one proton more and one neutron less or {\it vice versa}. The point to note is that this takes place {\bf within a barion family} of nuclei with a fixed number of nucleons, i. e. within the state space \$A\N\prod_{n=1}^{\otimes ,N_{nu}}\cH_n^{nu}\$ (cf (2)).

Mathematically, as known from textbooks on \QM describing spin, the operators \$t_+\equiv t_1+it_2\$ and \$t_-\equiv
t_1-it_2\$, the counterparts of  \$s_+\equiv s_x+is_y\$ and \$s_-\equiv
s_x-is_y\$, map a neutron state into a proton one (with the same spatial and spin sate) and {\it vice versa} (cf Preston 1962).

The transformation of a proton into a neutron or {\it vice versa} takes place within a nucleus. Quantum processes are described by unitary operators, which perform the change continually via the intermediate states that are superpositions of proton and neutron states. This is all very natural in the nucleon description and impossible in the proton-neutron one, where a so-called super-selection rule, prohibiting the mentioned intermediate states, is valid. 

In the epistemological scheme on how to improve our approximation of reality (see the Introduction) the last passages describe the 'more', the improvement that the nucleon theory yields in comparison with the proton-neutron theory. Hence, we can consider that it {\bf describes reality better}, i. e., that it is a better approximation to reality as far as the particles making up the nuclei are concerned.\\

\section{Answer to the anti-symmetrization dilemma for two identical fermions}

To begin with, let us consider a short {\bf historical} approach.

The inventor of the
exclusion principle, Pauli, is
reported to have said (private
communication by the late Rudolf E.
Peierls) that if two electrons are
apart, then they are distinct
particles by this very fact. His
principle applies to those that
are not in this relation.

Let us make possible a concrete discussion of fermions being apart.

\indent Let \$\cD_e\$ be a spatial
domain comprising the earth, and \$\cD_{out}\$ the complementary domain (in the set-theoretical sense, within all space). The two domains are,
of course, disjoint from each
other. The single-fermion {\bf distinguishing
projectors} are $$Q_i\equiv \int
\int \int_{\cD_i}\ket{\vec
r}\bra{\vec r} d^3\vec r,\enskip
i=\mbox{e,out}.\eqno{(10)}$$ They are orthogonal to each other due
to the disjointness of the
domains, and \$Q_e+Q_{out}=I\$, where \$I\$ is the identity operator.

Generalizing Pauli, Schiff (1955)
stipulates that two
{\it identical particles are
distinguishable} when the
two-particle probability amplitude
\$a(1,2)\$ of some dynamical
variable is different from zero
only when the two particles have
their values in {\it disjoint}
ranges of the spectrum of the
variable.

But, as De Muynck (1975)
remarks, this
actually {\it cannot ever occur}
when the wave function is
anti-symmetric, for then
\$a(2,1)=-a(1,2)\$.\\

Taking up Schiff's attempt to
formalize a generalization of
Pauli's distinguishing two identical
fermions, we assume that that Schiff's
two-particle amplitude \$a(1,2)\$
is a two-particle wave function. Then, we know from
textbooks that \$a(2,1)=-a(1,2)\$ for identical fermions.

Let, further, the index value in
\$Q^e_i,\$ and \$Q^{out}_i,\$ \$i=1,2\$ (cfb (10)) show to which of the two identical fermions the distinguishing
property applies. Then, {\it the
correct way to express Pauli's
criterion of distinguishability} is to
say that the two-particle system
{\it possesses} the property (cf (C.1) in Appendix C) expressed by the two-identical-fermion distinguishing projector
\$(Q^e_1Q^{out}_2+Q^{out}_1Q^e_2)\$:  $$(Q^e_1Q^{out}_2+Q^{out}_1Q^e_2)a(1,2)=
a(1,2)\eqno{(11)}$$ (cf (C.1)).

The general theory of distinguishing identical fermions expounded in Appendix B enables one to transform effectively the distinct extrinsic properties (being on earth or outside it in our concrete example) into
intrinsic ones by {\it isomorphic
transition} from the subspace
\$(Q^e_1Q^{out}_2+Q^{out}_1Q^e_2)A_{12}
(\cH_1\otimes\cH_2)\$
to the effective
distinct-particle state space
\$\Big(Q^e_1\cH_1 \otimes
Q^{out}_2\cH_2\Big).\$ Schiff's
mentioned criterion is actually
valid in the latter,
distinct-particle space.

Mirman's (1973) claim of
the essential role played by
experiments shows up in the fact
that the mentioned transformation
of extrinsic properties into
effective intrinsic ones is
restricted to experiments in which
the possession of the distinguishing property
\$(Q^e_1Q^{out}_2+Q^{out}_1Q^e_2)\$ is preserved.

Thus, a generalized Pauli
criterion of distinguishing
identical particles can be
expressed in the
quantum-mechanical formalism quite
satisfactorily as far as two
identical particles are concerned.

We see that Pauli's idea of two fermions being "apart", which is, no doubt, a spatial idea, should not be understood in the sense of distance (of 'far apart'), but only in the sense of disjoint domains.

Since disjoint-domain distinction is completely analogous to the proton-neutron difference, there is, obviously, no problem in {\bf extending the former distinction to any number of fermions}, in particular, to all fermions in the universe and to any number of disjoint domains, along the lines of the general 'distinct-identical-particle theory' of Appendix B. But there is an important difference in the two distinctions discussed (see subsection 4.1).

In all experiments done in
the laboratories on earth, the relevant observables satisfy the required restrictions
of compatibility with the corresponding
distinguishing property. But one must wonder if all important observables can be measured within  earthbound laboratories. (For a negative answer see subsection 4.1 .)\\

Anti-symmetrizing all fermions of a kind in the universe gives 'more' in princile (cf the Introduction) than Pauli's original cautious formulation for several reasons
with evident physical meaning:

(a) One assumes as little as possible (Occham's razor - the demand to economize in assumptions).

(b) The formalism does not favor space over other observables. Namely, it is clear that distinction in terms of disjoint domains in the spectrum of any other observable (or set of compatible observables) can take the place of the position observable. Thus, there is no need to find justification for the unique conceptual position of space in \qm .

Thus, according to the epistemological scheme advocated in the Introduction, universal anti-symmetrization for any kind of identical fermions is closer to reality then doing it in separate domains. In other words, we obtain a better theoretical approximation to reality in the described manner.\\

\section{Concluding remarks}

In this article a firm attitude is taken that there exists a \Q reality independent from the observer, and that we approach it with our theories like one approximates the irrationals on the real axis by rationals because the former can never be expressed exactly. The main point is that we can improve the approximation, i. e., make a better theory as explained in the Introduction. An example of such improvement is given in Appendix B for identical fermions. (It is a general theory how to distinguish identical fermions or bosons, but it is only a particular example as far as improving a theory is concerned.)

To make it more comprehensible, it is shown in some detail in section 2 that the concept of nucleons brings us closer to reality then the idea of protons and neutrons does. This concrete example of the identical-fermion theory in Appendix B has been called the nucleon paradigm because it is viewed as a basic example for the general scheme of making a better theory. Also the exclusion principle is discussed (in section 3) from the point of view of Appendix B and the scheme of how to improve a theory.\\

It is desirable to shed additional light on some salient features of the two cases of distinguishing identical fermions. We are interested in differences between some features of nucleons and and analogous features of fermions in disjoint spacial domains. We also want to have another look at the conversion of intrinsic into extrinsic particle properties, and {\it vice versa}, which makes the physical basis of the entire distinguishing theory.\\

Concerning the effective distinct-cluster description in Appendix B (we have two clusters both in the case of protons and neutrons and in our example of disjoint spatial domains), one should note that it is {\bf not an approximation} (as effective particles
often are). For states that possess the distinguishing property and for  observables that are compatible with it, the description is {\bf exact}, and for
those that do not possess it (are not compatible with it)  {\it it
does not make sense}.\\

We saw that Pauli himself mentioned 'being spatially apart' in the formulation of his principle. In case of the nuclear particles, his exclusion principle was also articulated separately for protons and separately for neutrons.

All this is not wrong, but it has turned out that one can do better, and thus make a theory that approximates  reality  more closely (see the Introduction).\\

\subsection{Differences}

The barion family discussed in section 2 consists of nuclei, and no superpositions of distinct nuclei in the same family are observed in nature. Since the many-nucleon distinguishing property requires precisely this, all nuclei possess the many-nucleon distinguishing properties with different number of protons \$N_{pr}\$.

This is not so in the case of spatial disjoint-domain distinction for some kind of fermions, e. g., nucleons. Particles can be, and often are, {\bf delocalized} spatially. If described by a wave function, it is a superposition of a component (wave function) that is in the domain of earth, and one that is outside. (Like in the case of passing a double slit, when the delocalized photons or massive particles that pass both slits simultaneously are the object of experimenting.)

Delocalized particles do not possess the many-particle distinguishing property, and hence they cannot be treated separately on earth, and separately outside earth. They must be omitted from the effective distinct-particle description. In this sense, the latter theory approximates the wider identical-fermion one even where the many-particle distinguishing property is observed. The fewer fermions are left out, the better the description.

One wonders if there is anything wrong
with applying \QM to a restricted domain, re. g., earth or a laboratory on earth. The answer is "yes". We give
an argument against the exactness of
local \QM of this kind.

When the orbital (or spatial)
tensor-factor space of a single
particle is determined by the
basic set of observables, which
are the position, the linear
momentum, and their functions, spin etc., one obtains an {\it irreducible}
space, i. e., a space that has no
non-trivial subspace {\it
invariant simultaneously} for all
the basic observables (for
position and linear momentum; cf
sections 5 and 6 in chapter VIII
of Messiah's (1961) book.
Hence, the above used subspace
\$Q_1^e\cH_1\$ (for the local,
earth quantum-mechanical
description) is not invariant
either. It is, of course,
invariant for position, but linear
momentum has to be replaced by
another Hermitian operator
{\it approximating} it.\\

\subsection{Converting
extrinsic properties into
intrinsic ones and {\it vice versa}}

\indent As it was stated, the notion of
identical particles rests on the
idea of equal intrinsic properties
of the particles. One can view the general theory expounded in Appendix B as the general
framework how to {\it convert some
extrinsic properties}, represented
by nontrivial projectors in the
single-particle state space, {\it
into intrinsic ones}. The
extrinsic properties are converted into intrinsic ones in terms of single-particle
distinguishing projectors
\$\{Q_j:j=1,2,\dots ,J\}\$ generating the distinct clusters (cf (B.1) and (B.2)). In the effective distinct-cluster
space \$\cH\N^D\$ (cf (B.3a-c)) these properties become actually intrinsic.\\

It is important to notice that the effective distinct-cluster
space \$\cH\N^D\$ is still expressed by projectors. But since the description is restricted to their ranges (one is within the space), they amount to the same a intrinsic properties.

One should also pay attention to the difference in our two two-cluster theories. In the nucleon paradigm one could even get rid of the single-nucleon distinguishing projectors \$Q^+\$ and \$Q^-\$ by eliminating the isospin tensor factor space in the single-nucleon space (1a) and going over to the proton space (1b) and the neutron space (1c). In the disjoint-spatial-domain distinction there is no suitable way to do something analogous. But when one restricts the description to the (invariant) range of the many-distinct fermion space \$\cH^D_{\N}\$, this {\it simulates} the convertion of the extrinsic property into an intrinsic one in a satisfactory manner.\\

Sometimes the {\it reverse
conversion} of intrinsic
properties into extrinsic ones
takes place. For this algorithm
the same conceptual framework from Appendix B can
be used. In other words, the theory presented in this article covers also this case.

The best example is that of
protons and neutrons, where historically
(and in many textbooks) the proton-neutron description is given priority if not presented exclusively.

The reverse process at issue consists
in transferring the quantum-mechanical
description from \$\cH\N^D\$ to the
subspace \$\cH\N^{id}\$ of the
first-principle space \$A\N\prod_{n=1}^{\otimes ,N_{nu}}   \cH_n\$.
Inclusion of \$\beta$-radioactivity
requires the use of the latter space
because that of the
former does not suffice.\\

Perhaps additional light is shed on the
{\it reverse} application if the
expounded theory by discussing a {\it
fictitious case}. Suppose we want to
treat the proton (pr) and the electron
(el) as two states of a single particle
(like the proton and the neutron). Can
we do this? The answer is affirmative,
and the way to do it is to use the
theory of this article in the, above
explained, reverse direction.

The new first-particle space would be
\$\cH_1\equiv Q_{pr}\cH_1\oplus
Q_{el}\cH_1,\$ where \$Q_{pr}\$ and \$Q_{el}\$
project \$\cH_1\$ onto the proton and
the electron subspace respectively. The
rest is analogous to the case of the
nucleons in section 2 with the important {\it
difference} that there is no
counterpart of the effect of the weak interaction. This means that every realistic
\$N$-particle state \$\rho\N^{id}\$
possesses the distinguishing property,
and can never lose it. Hence, the
corresponding distinct-cluster space
\$\cH\N^D\$ will always do for
description, and the simplicity requirement (the  razor of Occham) brings us back to
permanently distinct
particles.\\

Actually, one speaks of identical particles if
the particles have identical
{\bf complete} sets of intrinsic
properties.

This condition has the prerequisite
that long experience suggests that one
is unable to convert any of the
intrinsic properties by dynamical means
into extrinsic ones, and that one is
unable to extend the set of such
properties. These are {\it impotency
stipulations} analogous to those of
thermodynamics on which the
thermodynamical principles are based.

Let a good illustration be given for this. Some time ago
the electron neutrino and the muon
neutrino were believed to be identical
particles because they had their,
up-to-then known, intrinsic properties
in common. Later it was discovered that
they differ; the former has the
electronic leptonic quantum number, and
the latter the muonic one. Thus, their
other common properties were
incomplete; after completion it turned
out that they no longer have all
intrinsic properties equal.\\

An illustration for converting an
intrinsic property into an extrinsic
one is the case of parity and weak
interaction. Until the advent of the
famous parity-non-conserving weak
interaction experiments, parity could
be considered an intrinsic property of
the elementary particles. These
experiments converted it into an
extrinsic one, and nowadays we must
work with the parity observable with
its parity-plus and parity-minus
eigen-projectors.\\

{\bf Appendix A. The necessary textbook formalism and the anti-symmetrizer}
This Appendix has the purpose to remind the reader of the basic first-quantization (as distinct from second-quantization) textbook notions for the treatment of identical particles.\\

One has \$N\$ single-particle state spaces \$\{\cH_n:n=1,\dots ,N\}.\$ The
{\bf identicalness} of the particles is
expressed  (i) in terms of {\bf isomorphisms}
\$\{I_{m\rightarrow n}:m,n=1,\dots
,N;m\not= n\}\$ mapping the single-particle space \$\cH_m\$ onto \$\cH_n\$, \$m,n=1,2,\dots ,N\enskip m\not= n\$. Naturally, \$I_{m\rightarrow
n}I_{n\rightarrow m}=I_n,\$ \$I_n\$
being the identity operator in
\$\cH_n\$, \$n=1,2,\dots ,N\$.

Any two equally-dimensional Hilbert spaces are isomorphic, and there are very many different isomorphisms connecting them. For the identicalness the more important requirement is the following requirement on the isomorphism in (i): (ii) the physically meaningful operators, observables in particular, in each single-particle space are {\bf equivalent} with respect to the isomorphisms given in (i).

As an illustration, we
mention that the second-particle
radius-vector operator, e. g., is: \$\vec
r_2=I_{1\rightarrow 2} \vec r_1
I_{2\rightarrow 1}.$

The {\bf N-distinct-particle
space}, on which the description of
identical particles in
first-quantization \QM is based, is $$
\cH_{1\dots N}\equiv
\prod_{n=1}^{\otimes , N}\cH_n,\eqno{(A.1)}$$ where
\$\otimes\$ denotes the tensor (or
direct) product of Hilbert spaces. (We
shall use this symbol also for the
tensor product of vectors and of
operators.)\\

The {\it anti-symmetrizer}
(for identical fermions), written as a projector,  is $$
A_{1\dots N}\equiv (N!)^{-1}
\sum_{p\in \cS_N}(-1)^pP_{1\dots N}.\eqno{(A.2)}$$ Here \$\cS_N\$ is the so-called symmetric group, i. e., the group of all \$N!\$ permutations of \$N\$ objects, in this case of \$N\$ identical particles, by \$p\$ are denoted the elements of the group, \$(-1)^p\$ is the parity of the permutation. It is \$+1\$ if the permutation can be factorized into an even number of transpositions, and it is \$-1\$ of it can be factorized into an odd number of the latter (never can be both). The transpositions in \$\prod_{n=1}^{\otimes , N}\cH_n\$ are determined with the help of the isomorphisms \$\{I_{m\rightarrow n}:m,n=1,\dots
,N;m\not= n\}\$ defined above.

Finally, \$P_{1\dots N}\$ are the unitary operators that represent the permutations \$p\$ in the  N-distinct-particle space given by (A.1). When acting on an uncorrelated vector, \$P_{1\dots N}\$ permutes the tensor factor single-particle state vectors according to the prescription contained in \$p\$.

It should be noted that the {\bf state space of \$N\$ identical fermions} is $$A_{1\dots N}\cH\N\eqno{(A.3)}$$ (cf (A.1)).\\

{\bf Appendix B. How to obtain distinct
identical fermions}\\

We utilize the powerful tool of
projectors and elementary group theory.

{\it In the general case}, which we are
now going to elaborate, let the {\bf
distinguishing events} (or properties), which are going to generate the distinctness of the identical particles, be given by {\bf \$J\$ orthogonal
single-particle projectors}:
\$\{\{Q_n^j:j=1, \dots ,J\}:n=1,\dots
,N\},\$ \$\forall j, \forall n:\enskip
(Q_n^j)^{\dagger}=Q_n^j\$ (Hermitian
operators), \$\forall n:\enskip
Q_n^jQ_n^{j'}= \delta_{j,j'}Q_n^j\$
(orthogonal projectors), and finally,
\$\forall j:\enskip Q^j_n=
I_{1\rightarrow n}Q_1^jI_{n\rightarrow
1}, \enskip n=2,\dots ,N\$
(mathematically, equivalent projectors;
physically, same events or properties).

We have in mind {\bf \$J\$ clusters of
effectively-distinct particles},
\$2\leq J\leq N.\$We {\bf enumerate them by \$j\$} in an
{\it ordered} way according to the
(arbitrarily fixed) values of \$j:\$
\$j=1,\dots ,J.\$The \$j$-th cluster
contains a certain number of particles,
which we denote by \$N_j,\$ \$
\sum_{j=1}^JN_j=N.\$It will prove
useful to introduce also the sum of
particles up to the beginning of the
\$j$-th cluster: \$ M_j\equiv
 \sum_{j'=1}^{(j-1)}N_{j'}\$ for \$j\geq
 2,\$
 and \$M_1\equiv 0.\$

The single-particle distinguishing projectors appear in the distinct-particle space
\$\cH\N\$ (cf (A.1)) through a {\it tensor
product} determining the $N$-particle {\bf distinguishing projector} \$Q_{1\dots N}\$ in \$\cH\N\$:
$$Q_{1\dots N}\equiv
\prod_{j=1}^{\otimes ,J}
\Big(\prod_{n=(M_j+1)} ^{\otimes ,(M_j+N_j)}Q_n^j\Big).\eqno{(B.1)}$$ One
should note that the product in
the brackets applies to the \$j$-th
{\it cluster}, and  it consists of the tensor product of
 physically equal
(mathematically equivalent via
transpositions) single-particle
projectors (and there are \$J\$ clusters).

We want to introduce the corresponding
{\bf distinct-cluster space}
\$\cH_{1\dots N}^D,\$ which is the
state space consisting of the tensor product of \$J\$ {\bf distinct-particle clusters} (see (B.3a) and (B.3b) below) , each consisting of identical particles, and hence anti-symmetrized.

Let us call the {\bf 'cluster subgroup'}, and denote by \$S_N^{cl}\$, the subgroup of  permutations of \$N\$ objects that act possibly nontrivially only within the given clusters. (It is a subgroup of \$S_N\$, the group of all (N!) permutations of \$N\$ 0bjects.) The corresponding {\bf 'cluster anti-symmetrizer'}, we denote it by \$A\N^{cl}\$, is of the form $$A\N^{cl}\equiv\prod_{j=1}^{\otimes ,J}A_{(M_j+1)\dots (M_j+N_j)}=\sum_{p\in S_N^{cl}}(-)^pP\N\Big/\prod_{j=1}^J(N_j!),\eqno{(B.2)}$$

The distinct-cluster space
\$\cH_{1\dots N}^D\$ is defined as follows
$$\cH_{1\dots N}^D\equiv
\prod_{j=1}^{\otimes ,J}
\Big[A\M\Big(\prod_{n=(M_j+1)}
^{\otimes ,(M_j+N_j)}(Q^j_n\cH_n)\Big)\Big]=$$
$$\Big\{
\prod_{j=1}^{\otimes ,J} \Big[A\M
\Big(\prod_{n=(M_j+1)} ^{\otimes ,(M_j+N_j)}Q^j_n\Big)\Big]\Big\}\cH\N=$$
$$
Q_{1\dots N}A\N^{cl}\cH_{1\dots N},\eqno{(B.3a,b,c)}$$
where \$a,b,c\$ refer to the three obviously equivalent
expressions determining \$\cH_{1\dots
N}^D.\$ (and the two operator factors in
(B.3c) commute). Here by \$A\$ with indices running within one cluster are denoted the cluster anti-symmetrizers (cf (B.2)).

Note that the individual distinct {\it cluster
spaces} (factors in the tensor product
\$ \prod_{j=1} ^{\otimes ,J}\$ in (B.3a)) are {\it decoupled} from each
other (in the sense of
identical-fermion symmetry
correlations), i. e., one has the
tensor product \$\prod_{j=1}^{\otimes
,J},\$ but the factor spaces within each
cluster are coupled by the
corresponding anti-symmetrizers.

The order of the distinguishing projectors \$Q_n^j\$ in (B.1), within the clusters and of the clusters, is mathematically arbitrary and physically irrelevant. Hence, in view of the fact that we are dealing with identical fermions in \$A\N\cH\N\$, the suitable entity is not \$Q_{1\dots N}\$ given by (B.1). It is
the {\bf symmetrized distinguishing projector \$Q\N^{sym}\$} obtained from (B.1) by the permutation
operators:

$$Q_{1\dots N}^{sym}\equiv \bigg(
\sum_{p\in \cS_N} \Big(P_{1\dots
N}Q_{1\dots N}P_{1\dots
N}^{-1}\Big)\bigg)\Big/
\prod_{j=1}^J(N_j!).\eqno{(B.4)}$$ We
call it the {\bf distinguishing
property}, expressed as the symmetric projector (B.4) in \$\cH_{1\dots N}\$, which commutes with every permutation operator \$P\$. Hence, it commutes with the anti-symmetrizer \$A_{1\dots N}\$ (which is a linear combination of permutation operators cf (A.2)) and it reduces in the identical-fermion space \$A_{1\dots N}\cH_{1\dots N}\$.\\

Of central importance is the following
\$N$-identical-fermion subspace
\$\cH_{1\dots N}^{id}\$  of \$A_{1\dots N}\cH_{1\dots N}\$. It is
defined as the {\bf range of} \$Q\N^{sym}\$
in the identical-fermion state space \$A\N\cH\N\$:
$$\cH_{1\dots N}^{id}\equiv Q_{1\dots
N}^{sym}
A_{1\dots N} \cH_{1\dots N}=
A_{1\dots N}Q_{1\dots N}^{sym}
\cH_{1\dots N}
.\eqno{(B.5)}$$\\

{\bf\indent Theorem 1.} {\it The isomorphisms.} The subspaces
\$ \cH^{id}_{1\dots N}\$ and \$
\cH^D_{1\dots N}\$ of \$\cH\N\$ are {\it
isomorphic}, and the operators in \$\cH\N\$
$$I\N^{id\rightarrow D}\equiv
\Big((N!)\Big/\prod_{j=1}^J(N_j!)
\Big)^{1/2}Q_{1\dots N},
\eqno{(B.6)}$$
$$I\N^{D\rightarrow id}\equiv
\Big((N!)\Big/\prod_{j=1}^J(N_j!)
\Big)^{1/2} A_{1\dots
N}\eqno{(B.7)}$$ give rise to {\it
mutually inverse unitary isomorphisms}
mapping \$ \cH^{id}_{1\dots N}\$ onto
\$ \cH^D_{1\dots N}\$ and vice versa:
\$ \cH\N^D=I\N^{id\rightarrow
D}\cH\N^{id}\$ and \$ \cH\N^{id}=
I\N^{D\rightarrow id}\cH\N^D.\$\\

{\bf Proof.} {\it First}, we show that \$Q\N\$ maps \$\cH\N^{id}\$ into \$\cH\N^D\$.

One should note that \$Q\N^{sym}\$ is an orthogonal sum of \$N!\Big/\prod_{j=1}^JN_j!\$ (the number of cosets of the cluster subgroup \$S_N^{cl}\$ in \$S_N\$) projectors in \$\cH\N\$ (cf (B.1)). As a consequence, the subprojector relation \$Q\N \leq Q\N^{sym}\$ is valid: $$Q\N Q\N ^{sym}=Q\N .\eqno{(B.8)}$$

If \$P\N\$ is a permutation the action of which is {\it not restricted} to within the given clusters, then \$Q\N P\N\cH\N =Q\N (Q\N P\N\cH\N )=0\$ because in the subspace in the brackets at least one  single-particle distinguishing projector appears outside its cluster, and \$Q\N\$ (cf (B.1)) acting on it gives zero. Hence, in view of the definition of \$A\N\$ (cf (A.2)), and due to the fact that the 'cluster anti-symmetrizer' is of the form (B.2), one has $$Q\N A\N=(N!)^{-1}\prod_{j=1}^JN_j!Q\N A\N^{cl}.\eqno{(B.9)}$$

Then, on account of commutation of \$Q\N\$ with \$A\N^{cl}\$ (cf (B.1) and (B.2)) as well as (B.8), definition (B.3c) finally gives for every element in \$\ket{\Psi}\N\in\cH\N\$:   $$Q_{1\dots N}A\N Q_{1\dots N}^{sym} \ket{\Psi}\N=(N!)^{-1}\prod_{j=1}^JN_j!Q\N A\N^{cl}\ket{\Psi}\N\enskip\in\enskip\cH\N^D$$ (cf (B.3c)). Thus, \$I\N^{id\rightarrow D}\$ (cf (B.6)) maps \$\cH\N^{id}\$ into \$\cH\N^D\$ as claimed.\\

For the {\it proof in the opposite direction}, one should note that one also has the evident subprojector-relation \$A\N\leq A\N^{cl}\$: $$A\N A\N^{cl}=A\N.\eqno{(B.10)}$$ Hence, taking into account (B.9), we have for every element \$\ket{\Phi}\N\in\cH\N\$ (cf (B.3c)): $$A_{1\dots
N}Q\N A\N^{cl}\ket{\Phi}\N=$$ $$(N!)\Big(\prod_{j=1}^JN_j!\Big)^{-1}A\N\Big(A\N Q\N A\N)A\N^{cl}\ket{\Phi}\N.$$

On account of the relations $$\forall p\in S_N:\quad A\N P\N=P^{-1}\N A\N=(-)^pA\N ,\eqno{(B.11)}$$ which follow from the so-called 'translational invariance' of the group (in both directions) and the fact that taking the parity is a homomorphism, and in view of the definition (B.4), one obtains $$A\N Q\N^{sym} A\N =\Big(N!\Big/\prod_{j=1}^JN_j!\Big)
A\N Q\N A\N.\eqno{(B.12)}$$

Hence, one has further, due to (B.9), $$A_{1\dots
N}Q\N A\N^{cl}\ket{\Phi}\N=$$
$$\Big(N!\Big/\prod_{j=1}^JN_j!\Big)^{-1}
A_{1\dots N}Q\N A_{1\dots N}\ket{\Phi}\N=$$
$$Q\N^{sym} A\N\ket{\Phi}\N\enskip\in\enskip\cH\N^{id}$$ as claimed  (cf (B.3c) and (B.5)). The last step has take into account the commutation of \$A\N\$ with \$Q\N^{sym}\$ and the idempotency of the former.\\

{\it Next}, we show that the maps
\$I\N^{id\rightarrow D}\$  and
\$I\N^{D\rightarrow id}\$ in
application to the subspaces
\$\cH\N^{id}\$ and to \$\cH\N^D\$
respectively are each other's inverse.

Owing to the definitions (B.6) and (B.7),
and to the definition (A.2) of
\$A\N\$ one has the following
equality of maps:
$$I\N^{id\rightarrow D}I\N^{D\rightarrow
id}=\Big\{\Big((N!)\Big/\prod_{j=1}
^J(N_j!) \Big)^{1/2}Q_{1\dots N}
\Big((N!)\Big/\prod_{j=1}^J(N_j!)\Big)^{1/2}
A\N=$$
$$\Big\{\Big(\prod_{j=1}^J(N_j!)
\Big)^{-1}\sum_{p\in
\cS_N}(-)^pQ_{1\dots N} P_{1\dots
N}\Big\}.$$

For any \$\ket{\Phi}\N =Q\N A\N^{cl}\ket{\Phi}\N\in\cH\N^D\$ (cf (B.3c)), the definition (A.2) of the anti-symmetrizer implies $$I\N^{id\rightarrow D}I\N^{D\rightarrow
id}\ket{\Phi}\N =$$ $$\Big\{\Big(\prod_{j=1}^J(N_j!)
\Big)^{-1} \sum_{p\in
\cS_N}(-)^pQ_{1\dots N} P_{1\dots
N}Q_{1\dots N}\Big\}A\N^{cl}\ket{\Phi}\N=$$
$$\Big(\prod_{j=1}^J(N_j!)
\Big)^{-1} \sum_{p\in \cS_N}
Q_{1\dots N}\Big(P_{1\dots
N}Q_{1\dots N} P_{1\dots
N}^{-1}\Big)\Big((-)^pP_{1\dots
N}A\N^{cl}\Big)\ket{\Phi}\N.$$

All
\$\Big(P_{1\dots N}Q_{1\dots
N}P_{1\dots N}^{-1}\Big)\$ multiply
with \$Q_{1\dots N}\$ into zero {\it except when} \$ p\in S_N^{cl}\$, and then \$P\N Q\N P\N^{-1}=Q\N\$. Hence, the sum gives this projector \$\Big(\prod_{j=1}^J(N_j!)
\Big)\$ times. Besides, for \$p\in S_N^{cl}\$, \$P\N\$ commutes with \$A\N^{cl}\$ (cf (B.2)) and \$\Big((-)^pP_{1\dots
N}A\N^{cl}\Big)=A\N^{cl}\$ by an elementary argument analogous to that giving the adjoint of (B.11). Thus, finally, $$I\N^{id\rightarrow D}I\N^{D\rightarrow
id}\ket{\Phi}\N=Q\N A\N^{cl}\ket{\Phi}\N=\ket{\Phi}\N .$$
This establishes the claim that
\$I\N^{id\rightarrow D}\$ is the
inverse of \$I\N^{D\rightarrow id}$.\\

{\it Analogously}, in view of (B.6) and (B,7), we have the following equality
of operators due to (A.2):
$$I\N^{D\rightarrow id}I\N^{id\rightarrow
D}=\Big((N!)\Big/
\prod_{j=1}^J(N_j!)
\Big)^{1/2}A_{1\dots N}
\Big((N!)\Big/
\prod_{j=1}^J(N_j!)\Big)^{1/2}
 Q_{1\dots N}=$$
$$\Big(\prod_{j=1}^J(N_j!)\Big)^{-1}
\sum_{p\in \cS_N}\Big(P_{1\dots N}
Q_{1\dots N}P_{1\dots
N}^{-1}\Big)(-)^p P_{1\dots
N}.$$ for any \$\ket{\Psi}\N =A\N Q\N^{sym}\ket{\Psi}\N\in\cH\N^{id}\$, one can write $$I\N^{D\rightarrow id}I\N^{id\rightarrow
D}\ket{\Psi}\N =$$  $$
\Big(\prod_{j=1}^J(N_j!)
\Big)^{-1}\sum_{p\in
\cS_N}\Big(P_{1\dots N} Q_{1\dots
N}P_{1\dots N}^{-1}\Big)(-)^p
P_{1\dots
N}A\N\ket{\Psi}\N =$$
$$Q\N^{sym}A\N\ket{\Psi}\N .$$ In the last step we have utilized the adjoint of (B.11) and the definition (B.4). Thus,
the claim that the two maps are the
inverse of each other is proved.

Since the maps are the inverse of each
other, it is easily seen hat they are
necessarily surjections and injections,
i. e., bijections as claimed.\\

{\it Next}, we prove that \$I\N^{D\rightarrow
id}\$ preserves the scalar product,
which we write as \$\Big(\dots ,\dots
\Big).\$Let \$\Psi\N\$ and \$\Phi\N\$
be two arbitrary elements of
\$\cH_{1\dots N}^D.\$On account of (B.7) one has
$$\Big(I\N^{D\rightarrow Id}\Psi\N
,I\N^{D\rightarrow Id}\Phi\N\Big)=$$
$$\6\Big(\Psi\N ,A\N\Phi\N\Big).$$
Further, on account of the
fact that both \$Q\N\$ and \$A\N^{cl}\$ acts as the identity
operator on \$\Psi\N\$ and \$\Phi\N\$ (cf (B.3c)), one can write
$$lhs=\Big(\prod_{j=1}^J(N_j!)\Big)^{-1}
\sum_{p\in \cS_N}\Big(\Psi\N
,Q\N (P\N Q\N P\N ^{-1})\Big((-)^pP\N A\N^{cl}\Phi\N\Big).$$ Again
one has \$ Q\N (P\N Q\N
P\N ^{-1})=0,\$ except if \$ p\in
S_N^{cl}\$, when it is equal to \$Q\N.\$
 Therefore, $$lhs=\Big(\Psi\N
,Q\N A\N^{cl}\Phi\N\Big)= \Big(\Psi\N
,\Phi\N\Big)$$ as claimed.\\

It is easy to see that also the inverse
of a scalar-product preserving
bijection must be scalar-product
preserving. {\it This completes the
proof of Theorem 1.}\\

The physical meaning of the identical-fermion symmetry {\it
decoupling} and the {\it coupling
isomorphisms} \$I\N^{id\rightarrow D}\$
and \$I\N^{D\rightarrow id}\$
respectively given in the theorem shows
up primarily, of course, in the {\it observables}
that are defined in \$\cH\N ^{id}\$ and
\$\cH\N ^D.\$ The corresponding or
equivalent operators (obtained by the
similarity transformation) are of the
same kind: Hermitian, unitary,
projectors etc. because all these
notions are defined in terms of the
Hilbert-space structure, which is
preserved by the (unitary)
isomorphisms.\\

It is seen that a prerequisite for
describing an evolution or a
measurement in the subspaces \$\cH\N
^{id}\$ and \$\cH\N ^D\$ is the
possession of the distingwishing
properties (occurrence of the events)
\$Q\N^{sym}\$ and \$Q\N\$ respectively,
and their preservation.

A relevant
observable for the decoupling, i. e., a
Hermitian operator that reduces in
\$\cH\N ^{id},\$ is one that {\it
commutes} with the distinguishing
projector \$Q\N ^{sym},\$ and {\it one
confines oneself to its reducee} in
\$\cH\N^{id}\$. In physical
terms, the observable must be {\it
compatible} with the distinguishing
property \$Q\N ^{sym}\$ and one must
assume that the property is {\it
possessed} (cf (C.1) and (C.2) in
Appendix C), and that this is preserved
if some process is at issue.\\

{\it\indent Theorem 2. A)} Let
\$O_{1\dots N}^D\$ be an operator in \$\cH_{1\dots N}\$ (with a physical meaning) that {\bf commutes} (is compatible) {\bf both}
with every permutation operator permuting possibly non-trivially only within each cluster ('cluster permutations')
and with \$Q_{1\dots N}\$ (cf (B.1)). (Hence,
\$O_{1\dots N}^D\$ reduces in \$\cH_{1\dots N}^D\$, cf (B.3c).) Let,
further,
$$O_{1\dots N}^{id,(D)}\equiv
\Big(\prod_{j=1}^J(N_j!)\Big)^{-1}
\sum_{p\in\cS_N}P_{1\dots N}O_{1\dots N}^DQ\N P_{1\dots N}^{-1}\eqno{(B.13)}$$ be the symmetrized
product of \$O_{1\dots N}^D\$ and \$Q\N\$. Then \$O_{1\dots N}^{id,(D)}\$
commutes with every permutation operator (it is a 'symmetric' operator), hence with \$A\N\$ (cf (A.2)), and with
\$Q_{1\dots N}^{sym}\$ (cf (B.4)),
and the {\it reducee} of \$O_{1\dots N}^D\$ in
\$\cH_{1\dots N}^D\$ (cf (B.3c)) and that of
\$O_{1\dots N}^{id,(D)}\$ in \$\cH_{1\dots N}^{id}\$ (cf (B.5))
respectively are {\bf equivalent}
(physically the same observables) with
respect to the isomorphisms in Theorem
1. One can express this in \$\cH\N\$ by the operator equality:
$$O_{1\dots N}^{id,(D)}A\N Q\N^{sym}=
\Big(I_{1\dots N}^{D\rightarrow id}\Big)O_{1\dots N}^D
\Big(I_{1\dots N}^{id\rightarrow D}\Big)A\N Q\N^{sym}
\eqno{(B.14a)}$$ (cf (B.5)).

{\it B)} Conversely, let \$B_{1\dots N}^{id}\$ be a symmetric operator in \$\cH_{1\dots N}\$ (with a physical meaning) that {\bf
commutes} (is compatible)  with
\$Q_{1\dots N}^{sym}\$. Then the
operator
$$B_{1\dots N}^{D,(id)}\equiv Q_{1\dots N}
B_{1\dots N}^{id}Q_{1\dots N} \eqno{(B.14b)}$$ (in \$\cH_{1\dots N}\$)
commutes with every cluster
permutation, hence with \$A\N^{cl}\$ (cf (B.2)), and with \$Q_{1\dots N}\$ (cf (B.1)). The reducee of \$B_{1\dots N}^{D,(id)}\$ in \$\cH_{1\dots N}^D\$
is {\rm equivalent} with (physically
the same observable as) the reducee of
\$B_{1\dots N}^{id}\$ in \$\cH_{1\dots N}^{id}.\$ The equivalence is given by the operator relation in \$\cH\N\$:
$$B_{1\dots N}^{D,(id)}A\N^{cl}Q\N=\Big(I_{1\dots N}^{id
\rightarrow D}\Big)
B_{1\dots N}^{id}\Big(I_{1\dots N}^{D\rightarrow
id}\Big)A\N^{cl}Q\N
\eqno{(B.14c)}$$ (cf (B.3c)).\\

{\it Proof. A)} Since \$\forall p'\in S_N\$, also \$\{P\N'P\N:\forall p\in S_N\}\$ is the symmetric group (so-called translational invariance of groups), one has $$\forall p'\in S_N:\quad P'\N O_{1\dots N}^{id,(D)}(P'\N)^{-1}=$$ $$
\Big(\prod_{j=1}^J(N_j!)\Big)^{-1}
\sum_{p\in\cS_N}(P'\N P_{1\dots N})O_{1\dots N}^D(P'\N P_{1\dots N})^{-1}=O_{1\dots N}^{id,(D)},$$ which is obviously equivalent to commutation of the operator with every permutation operator. Hence it commutes also with \$A\N^{cl}\$ (cf (B.2)).

On account of the facts that both \$Q\N\$ and \$O\N^D\$ commute with every cluster permutation, we choose arbitrarily one permutation \$p_k\$ from each {\it coset} of \$S_N^{cl}\$ in \$S_N\$, i. e., we view \$S_N\$ as the set-theoretical sum of cosets \$S_N=\sum_{k=1}^Kp_kS_N^{cl}\$, where \$K=N!\Big/\prod_{j=1}^JN_j!\$. Then we can write $$Q\N^{sym}=\sum_{k=1}^KP\N^kQ\N (P\N^k)^{-1}\eqno{(15)}$$ (cf (B.4)), and $$O\N^{id,(D)}=\sum_{k=1}^KP\N^kQ\N O\N^DQ\N (P\N^k)^{-1}\eqno{(16)}$$ (cf (B.13)). Further, $$(P\N^kQ\N (P\N^k)^{-1})(P\N^{k'}Q\N (P\N^{k'})^{-1}=$$  $$(P\N^{k'}Q\N (P\N^{k'})^{-1})(P\N^kQ\N (P\N^k)^{-1}=0\quad\mbox{if}\quad k\not= k'$$ (cf (B.1)). Therefore, $$Q\N^{sym}O\N^{id,(D)}=O\N^{id,(D)}Q\N^{sym}=
O\N^{id,(D)}.$$ We thus have commutation of \$O\N^{id,(D)}\$ with \$Q\N^{sym}\$, and the former operatorn reduces in \$\cH\N^{id}\$ (cf (B.5)).

The claimed relation (14a) is explicitly: $$O_{1\dots N}^{id,(D)}A\N Q\N^{sym}=\Big[(N!)\Big/\Big(\prod_{j=1}
^J(N_j!)\Big)\Big]A\N O\N^DQ\N A\N Q\N^{sym}.\eqno{(B.17)}$$ In analogy with \$(B.12)\$, one has $$A\N O\N^DQ\N A\N=\Big[(N!)\Big/\Big(\prod_{j=1}
^J(N_j!)\Big)\Big]^{-1}A\N O\N^{id,(D)}A\N.$$ Replacing this in rhs(B.17), one obtains $$rhs(B.17)=
A\N O\N^{id,(D)}A\N Q\N^{sym}.$$ Finally, on account of commutation of the symmetric operator \$O\N^{id,(D)}\$ with the (idempotent) projector  \$A\N\$ (cf (A.2)), \$rhs(B.17)=lhs(B.17)\$.\\

{\it B)} Since \$B\N^{id}\$ is by assumption a symmetric operator and \$Q\N\$ commutes with each cluster permutation operator (cf (B.1)), so does \$B\N^{D,(id)}\$. Hence, \$B\N^{D,(id)}\$ commutes also with \$A\N^{cl}\$ (cf B.2)). The former operator obviously commutes with the (idempotent) projector \$Q\N$. Therefore it reduces in \$\cH\N^D\$ (cf (B.3c)).

In view of (B.14b), the claimed relation (B.14c) has the following explicit form $$Q_{1\dots N}
B_{1\dots N}^{id}Q_{1\dots N} A\N^{cl}Q\N=$$ $$\Big[(N!)\Big/\Big(\prod_{j=1}
^J(N_j!)\Big)\Big]Q\N
B_{1\dots N}^{id}A\N A\N^{cl}Q\N.
\eqno{(B.18)}$$ One can see that the rhs indeed equals the lhs due to the commutation of the projectors \$Q\N\$ and \$A\N^{cl}\$ and on account of the adjoint of (B.9). {\it This ends the proof.}\\

In case the
state (density operator) \$\rho\N^{id}\$ of an
\$N$-identical-fermion system
satisfies the relation
$$Q\N^{sym}\rho\N^{id}
=\rho\N^{id} ,\eqno{(B.19)}$$ we say that
the system {\bf possesses} the {\it
distinguishing property}
\$Q^{sym}_{\N}\$ in the state in
question (cf (C.2) in Appendix C). In this case, and only
in this case, it is amenable to Theorem
2.\\

It is important to notice that the theory presented in this Appendix is applicable to identical bosons equally as to identical fermions, one muist only replace the anti-symmetrizer projector \$A\N\$ (cf (A.2)) by the symmetrizer projector \$S\N\equiv\sum_{p\in S_N}P\N\Big/N!\$.

This theory was presented in more detail and in a form valid simultaneously for fermions and bosons in (Herbut 2006). Also much relevant mathematical help was included, especially about the symmetric group. But reading it, there is a price to pay: the exposition is less readable.\\


{\bf Appendix C. Possessed property or event that has occurred}\\

Let \$E\$ be a projector (physical meaning: property or event).

{\bf A) Pure-state case.}
Let \$\ket{\psi}\$ be a state vector (pure state). Then $$\bra{\psi}E\ket{\psi}=1\quad
\Leftrightarrow\quad E\ket{\psi} =\ket{\psi},\eqno{(C.1)}$$ where "$\Leftrightarrow$" denotes logical implication in both directions.

{\it Proof.} $$\bra{\psi}E\ket{\psi}=1\quad
\Leftrightarrow \bra{\psi}E^{\perp}\ket{\psi})=0,$$ where \$E^{\perp}\equiv I-E\$, \$I\$ being the identity operator, and \$E^{\perp}\$ is the ortho-complementary projector. Further,
$$\bra{\psi}E\ket{\psi}=1\quad\Leftrightarrow\quad ||E^{\perp}\ket{\psi}||=0\enskip\Leftrightarrow\enskip E^{\perp}\ket{\psi}=0\enskip\Leftrightarrow\enskip E\ket{\psi} =\ket{\psi}.$$\hfill $\Box$\\

{\bf B) General-state case.} Let \$\rho\$ be a density operator (general state). Then $$\tr(E\rho )=1 \quad\Leftrightarrow\quad E\rho =\rho.\eqno{(C.2)}$$

{\it Proof.} Let \$\rho =\sum_ir_i\ket{i}\bra{i}\$ be a spectral form of \$\rho\$ in terms of its eigen-vectors \$\{\ket{i}:\forall i\}\$ corresponding to its positive eigenvalues \$\{r_i:\forall i\}\$. (It always exists.) Then
$$\tr(E\rho )=1\quad\Leftrightarrow\quad
\sum_ir_i\tr(E\ket{i}\bra{i})=1 \quad\Leftrightarrow\quad\sum_ir_i\bra{i}E\ket{i}=1
\quad\Leftrightarrow\quad \$$ $$\sum_ir_i(1-\bra{i}E\ket{i})=0 \quad\Leftrightarrow\quad\forall i:\enskip \bra{i}E\ket{i}=1$$  $$
\quad\Leftrightarrow\quad\forall i:\enskip E\ket{i}=\ket{i}\quad \Leftrightarrow \quad\forall i:\enskip E\rho =\rho.$$ In the last but one step (C.1) has been made use of.\hfill $\Box$\\

\noindent
{\bf References}\\

\noindent
Adler, C. G. (1989). "Realism and/or
physics". {\it American Journal of Physics.}

{\bf 57} 878-882.

\noindent
De Muynck, W. M. (1975). Distinguishable- and indistinguishable- parti-

cle descriptions of systems of identical particles. {\it International Journal of

Theoretical Physics,} {\bf 14}, 327-346.

\noindent
Herbut, F. (2001). How to distinguish identical particles. {\it American Journal

of Physics.} {\bf 69}, 207-217.

\noindent
Herbut, F. (2006). How to
distinguish identical particles. The general case.

ArXiv:quant-ph/0611049v1

\noindent
Jauch, J. M. (1966).
{\it Foundations of Quantum Mechanics.}
Addison-Wesley,

Reading,
Massachusetts. Chap. 15.

\noindent
Messiah, A. (1961). {\it
Quantum Mechanics.} North-Holland,
Amsterdam.

\noindent
Mirman, R. (1973). Experimental meaning of the concept of identical parti-

cles.
{\it Il Nuovo Cimento}, {\bf 18B} 110-111.

\noindent
Preston, M. A. (1962). {\it Physics of the nucleus.} Addison-Wesley, Reading,

Massachusetts.

\noindent
Schiff, L. I. (1955). {\it
Quantum Mechanics.} McGraw-Hill
Inc., New York,

Chap. 9.

\end{document}